\documentclass{egpubl}
 
\JournalPaper         
\usepackage[T1]{fontenc}
\usepackage{dfadobe}  

\usepackage{cite}  
\BibtexOrBiblatex
\electronicVersion
\PrintedOrElectronic
\ifpdf \usepackage[pdftex]{graphicx} \pdfcompresslevel=9
\else \usepackage[dvips]{graphicx} \fi

\usepackage{egweblnk} 

\title[BI-LAVA]%
      {BI-LAVA: Biocuration with Hierarchical Image Labeling through Active Learning and Visual Analysis}

\author[J. Trelles et al.]
{\parbox{\textwidth}{\centering Juan Trelles$^{1}$\orcid{0000-0002-0367-9763}
        and Andrew Wentzel$^{1}$\orcid{0000-0002-2003-2750} 
        and William  Berrios$^{2}$ 
        and Hagit Shatkay$^{3}$
        and G. Elisabeta Marai$^{1}$\orcid{0000-0002-7212-9669} 
        }
        \\
{\parbox{\textwidth}{\centering $^1$University of Illinois Chicago, USA\\
         $^2$Universidad Nacional de Ingenieria, Peru\\
        $^3$ University of Delaware, USA
       }
}
}

\begin{document}


\maketitle
\begin{abstract}
   In the biomedical domain, taxonomies organize the acquisition modalities of scientific images in hierarchical structures. Such taxonomies leverage large sets of correct image labels and provide essential information about the importance of a scientific publication, which could then be used in biocuration tasks. However, the hierarchical nature of the labels, the overhead of processing images, the absence or incompleteness of labeled data, and the expertise required to label this type of data impede the creation of useful datasets for biocuration. From a multi-year collaboration with biocurators and text-mining researchers, we derive an iterative visual analytics and active learning strategy to address these challenges. We implement this strategy in a system called BI-LAVA—Biocuration with Hierarchical Image Labeling through Active Learning and Visual Analysis. BI-LAVA leverages a small set of image labels, a hierarchical set of image classifiers, and active learning to help model builders deal with incomplete ground-truth labels, target a hierarchical taxonomy of image modalities, and classify a large pool of unlabeled images. BI-LAVA’s front end uses custom encodings to represent data distributions, taxonomies, image projections, and neighborhoods of image thumbnails, which help model builders explore an unfamiliar image dataset and taxonomy and correct and generate labels. An evaluation with machine learning practitioners shows that our mixed human-machine approach successfully supports domain experts in understanding the characteristics of classes within the taxonomy, as well as validating and improving data quality in labeled and unlabeled collections.


\end{abstract}  
\section{Introduction}

Labeled image datasets are required in multiple practical applications of supervised machine learning (ML), from autonomous driving~\cite{gou2020vatld}, spambot detection~\cite{Khayat2020VASSL} and medicine~\cite{sager2021survey, baidakova2021medium} to the image-based retrieval and identification of biomedical publications in biocuration~\cite{trelles2021animo, trelles2023enhanced,trabucco2020modality}. Not surprisingly, labeling solutions are urgently needed, rapidly creating a big industry market forecasted to generate a revenue of more than \$17 billion by 2030~\cite{gvr2023report}. The most common labeling settings involve recognizing and labeling the content captured in photographs, which is often a task humans excel at, even without training (e.g., recognizing the content as a boy holding a toothbrush), and which the industry rewards with relative low-pay (\$1 to \$25 per task)~\cite{andrewng2023}. However, other labeling tasks, for example aimed at identifying a specialized image type,  require advanced knowledge at the post-graduate level, and would result in a pay in excess of \$300 per task. For example, biocuration aims to organize into taxonomies and ontologies the vast information published in the biomed field. Although biomed scientific publications provide a vast source of unlabeled images, identifying the exact image type and subtypes from such a taxonomy (e.g., Experimental --> Gel --> Northern blot) can typically be performed only by highly skilled individuals named biocurators. 

In such settings, incremental ML strategies, including but not limited to active learning (AL)~\cite{bernard2018vial}, can help reduce the labeling costs by identifying smaller image subsets that require human intervention~\cite{settles2009active} and then learning from these smaller sets. In addition, strategies can leverage high-confidence predictions on the unlabeled set to add more labels, even though they do not come from a human source—i.e., they are pseudo- or weak-labels. Finally, integrating such ML strategies with visual analytics (VA) can also facilitate identifying and selecting labeling candidates from unlabeled data pools, and discovering knowledge in unfamiliar datasets.

When it comes to practical applications like the curation of biomedical data, several challenges from model and user perspectives hinder the integration of VA and ML techniques for understanding and labeling images into classes. From a model perspective, unlike typical systems~\cite{liu2018interactive, xiang2019interactive, chen2018anchorviz, bauerle2020classifier, chen2021interactive} that handle a single classifier and a single, non-hierarchical classification scheme (i.e., flat taxonomy), biocuration fundamentally relies on hierarchical taxonomies that may require multiple classifiers. This domain requirement complicates both the ML solution and the VA approach, because available labeled data may not match the taxonomy at the required level of specification—e.g., an image labeled as \textit{microscopy} does not indicate whether it belongs to the \textit{light} or \textit{fluorescence microscopy} subcategories, i.e., the provided label is incomplete. Further issues in data quality, such as mislabels, imbalanced distributions of classes, and a lack of representative samples, can affect the model's performance~\cite{yuan2020survey}. In other words, the ML + VA solution should be able to handle incomplete ground-truth labels for many image sets, at multiple hierarchical levels, and possibly multiple classifiers.

From the user perspective, non-domain experts such as model builders may lack the expertise to correct and label domain-specific images. For example, classifying traffic light pictures is more manageable than distinguishing between \textit{gel} and \textit{plate} figures. Furthermore, although popular labeling tools such as Label Studio~\cite{LabelStudio} or napari~\cite{napari} can process a single image at a time, they lack the features necessary to process image collections, as in biocuration. Overall, model builders spend more time working with data than with models~\cite{anaconda2022survey}, yet they have limited tools to understand unfamiliar image collections. A VA solution should support understanding such collections, because this type of support is crucial for labeling data efficiently and improving models' performance.

In this paper, we introduce BI-LAVA (\textbf{B}iocuration with Hierarchical \textbf{I}mage \textbf{L}abeling through \textbf{A}ctive \textbf{L}earning and \textbf{V}isual \textbf{A}nalysis), a system that integrates feedback from an incremental ML strategy to aid the hierarchical exploration, understanding, and labeling of an unfamiliar image dataset. BI-LAVA is the first system to support novice labelers working with unfamiliar datasets. BI-LAVA is the only system to date to handle hierarchical classification schemes. BI-LAVA scales to thousands of images, and in that process, deals with cluttering issues via spiral image layouts which capture the neighborhood projections from embedding space to 2D. Unlike any other labeling system, BI-LAVA’s ML backbone enables further feedback to help labelers understand problems with the classification models in terms of the data used for training (including incomplete labels), testing, or validation, and low and high-confidence samples. 

Beyond the specific system capabilities above: 1) We introduce, characterize and document the biomedical image labeling process, from the novel perspective of model builders who collaborate with domain experts in biocuration.  2) We design and build a novel VA labeling system (BI-LAVA) to support this process, and leverage custom encodings and interactions. 3) We report and discuss feedback and lessons learned from a BI-LAVA evaluation with ML practitioners working with an unfamiliar collection of biomedical images.

\section{Related work}

\subsection{AL and visualization for image classification}

While our ML backbone could leverage a number of different learning strategies, as long as they can identify low and high-probability samples, our current instantiation uses AL and Cost-Effective Active Learning (CEAL)~\cite{wang2017ceal}. CEAL considers samples with low entropy and high probability of being correctly classified as pseudo-labels which are used to train the model in future steps. Low-confidence, high entropy labels need to be acted on by human labelers. Other approaches include Core-Set~\cite{sener2018active}, which looks for the best subset for AL. However, Core-set is not suitable for large-scale labeled datasets. BI-LAVA uses CEAL due to its inexpensive, simple approach for separating low and high probability predictions. CEAL also produced promising preliminary results on our biomedical dataset, although it could be easily replaced by alternative approaches.

Previous work on Visual Interactive Labeling (VIAL) has explored the integration of visualization to augment human-in-the-loop AL,~\cite{bernard2018vial, sacha2014knowledge, bernard2018comparing}. Other work has identified a taxonomy of data types and tasks in VIAL~\cite{Bernard2021Taxonomy} with single views. We expand on this by exploring tasks for multiple linked views. Other works have explored multi-instance labeling with self-organizing maps~\cite{moehrmann2010improving} and pixel averages~\cite{hoferlin2011inter} to speed up labeling. In contrast, we focus on exploring and labeling image neighborhoods around an image of interest.

Beyond AL, visualization research can generally aid in labeling image data. Several visualization systems attempt to detect and correct labeling errors~\cite{liu2018interactive, xiang2019interactive, bauerle2020classifier}, identify a lack of representation in the dataset~\cite{chen2018anchorviz}, and leverage unlabeled data~\cite{chen2021interactive}. Visual feedback also enables understanding model performance~\cite{rauber2018projections, Bernard2021Taxonomy}. Our work considers a more complex combination of these goals. Specifically, we target correcting label errors and analyzing labeled and unlabeled images, and we provide visual support for understanding data collections and model behavior. Furthermore, although most of these works employ AL, none deal with hierarchical taxonomies connected to multiple classifiers.

Two other works deal with related problems, but not for image data.  VIANA~\cite{sperrle2019viana} enables annotating argumentation data using scrollable interactions to transition between different levels of aggregation, similar to our hierarchical taxonomy. However, unlike VIANA, we must always display the image labels and rely on a multiple-view system for annotating data and encouraging familiarization with the dataset. A second system, VASSL~\cite{Khayat2020VASSL}, uses multiple views to identify spambots by focusing exploration around an item of interest and filtering over dimensionally reduced elements. However, VASSL focuses on binary labeling. Still, both VIANA and VASSL are good examples of the need for contextual data for labeling, a current limitation in labeling systems like Label Studio~\cite{LabelStudio} or napari~\cite{napari}.


\subsection{Exploring sets or collections of images}

Dimensionality reduction algorithms enable viewing and exploring large unstructured sets and collections of images by transforming features derived from colors or latent representations to two dimensions.  Markers and thumbnails on scatterplots are commonly used to display images~\cite{nguyen2008interactive}, but they quickly clutter due to limits of display sizes. Alternatives to deal with cluttering include hierarchical clustering with zooming~\cite{chen2020oodanalyzer, xiang2019interactive}, scatterplots supported by grid-layouts to display thumbnails~\cite{xiang2019interactive, liu2018interactive, chen2021interactive}, or density contours to summarize data distributions and display a few samples~\cite{mayorga2013splatterplot}. Our design combines techniques from density contours with filters, selections, and supporting views to organize thumbnails.

Complementary techniques attempt to organize thumbnails to minimize clutter. Grid-layout algorithms transform image positions from a projected space to a grid layout while preserving neighborhood similarities~\cite{Cutura2021Hagrid, Fried2015IsoMatch, Duarte2014Nmap, bertucci2022dendropmap,Gomez2016Dealing}. While static treemaps may hurt thumbnail visibility~\cite{Ghoniem2015WM}, proposed variations support image exploration enabling interactions~\cite{bertucci2022dendropmap} or inspecting local neighborhoods around an image of interest~\cite{wang2015similarity}. Further techniques allow repositioning image projections~\cite{abuthawabeh2020force} and summarizing groups~\cite{lekschas2020generic} but are limited to exploring a small subset. As BI-LAVA follows the rationale of exploring images of interest based on static positions, we draw inspiration from spiral treemaps layouts~\cite{wang2015similarity} for displaying thumbnails.


\subsection{Hierarchy visualizations}
Taxonomies of image modalities follow a hierarchical structure with different depths per branch. Typical encodings for this type of hierarchical data include indented lists~\cite{dudavs2018ontology}, node-link diagrams (e.g., trees), icicles~\cite{kruskal1983icicle,van2020space} and treemaps~\cite{shneiderman1998treemaps}. However, juxtaposed encodings are required to add further details~\cite{vehlow2015state}. To aim for a compact representation of our taxonomy and distribution of samples, we use a combination of indented lists and horizontal bar charts, similar to work done for confusion matrices~\cite{gortler2021neo}.


\begin{figure}[htb]
  \centering  
  \includegraphics[width=\linewidth]{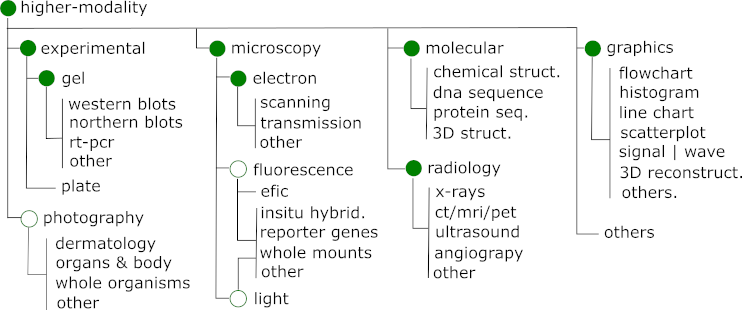}
  \caption{\label{fig:taxonomy}
           Taxonomy of image modalities for biomedical images in COVID-19 papers~\cite{trelles2021animo}. Nodes indicate parent classes with image classifiers; filled nodes further denote available classifiers at the start of the project. Nodes with blank fills denote expected classifiers where we did not have labeled data. For \textit{gels}, \textit{light} and \textit{fluorescence}, sub-classes denote experimental methods.}
\end{figure}

\section{Application background}
\label{section:background}
Unlike typical data labeling problems, we describe the challenge of unfamiliar domains, the issue of hierarchical incomplete labels, and specific tasks related to the intrinsic need for hierarchical taxonomies of labels and multiple classifiers.

Biocurators extract knowledge from biomedical publications to populate scientific databases, saving an immeasurable amount of time to fellow researchers~\cite{isb2018biocuration}. While text-based ML models accelerate biocuration tasks, such as document classification~\cite{jiang2020integrating, burns2019building}, work from our collaborators found that complementing these models with image acquisition modality information produces better results~\cite{li2021utilizing, shatkay2006integrating}. Such results are consistent with the claims of image importance in communicating scientific content~\cite{yu2009sufficient, shatkay2006integrating,cohen2003understanding}. However, integrating these modalities is underexplored due to the lack of standardization of classification schemes and the consequent lack of labeled data to train supporting image classifiers. 

Modern attempts to automatically classify biomedical images into taxonomies use deep learning models~\cite{GSB2016}. However, these taxonomies are not detailed at the level of granularity desired in biocuration. For example, they often classify images that use microscopy, but do not consider the modality (e.g., electron vs light microscopy) or sub-modality (e.g., scanning vs transmission microscopy), which has important information on the kind of experiments performed. To support creating more granular labels from sources with few or incomplete labels, we developed a hierarchical set of image classifiers following the parent nodes of a taxonomy developed for biocuration (Fig.~\ref{fig:taxonomy}).

Over the last five years, we have collaborated remotely with biocurators and text-mining researchers at three sites (Caltech, Jackson Labs, and Delaware) to integrate image and textual features to aid biocuration. Our team comprises two visual computing researchers and a senior undergraduate in ML. Some of our goals include harvesting labels for biomedical images and training image classifiers for image modalities, which lead to us exploring approaches for improving data labeling and the identification of modalities relevant to biocuration~\cite{trelles2021animo}. Through a series of semi-structured interviews, repeated observation, and feedback meetings, we became familiar with the existing labeling workflow and the image modality taxonomy to be used for classification.


\section{Methods}
Unlike other systems, our solution leverages multiple views of images and supports the multiple classifiers required by hierarchical taxonomies. BI-LAVA furthermore supports exploring unfamiliar hierarchical datasets to understand model behaviors, identify label errors, evaluate the ML performance, and identify data needs.

\begin{figure*}[htb]
  \centering
  \includegraphics[width=0.8\linewidth]{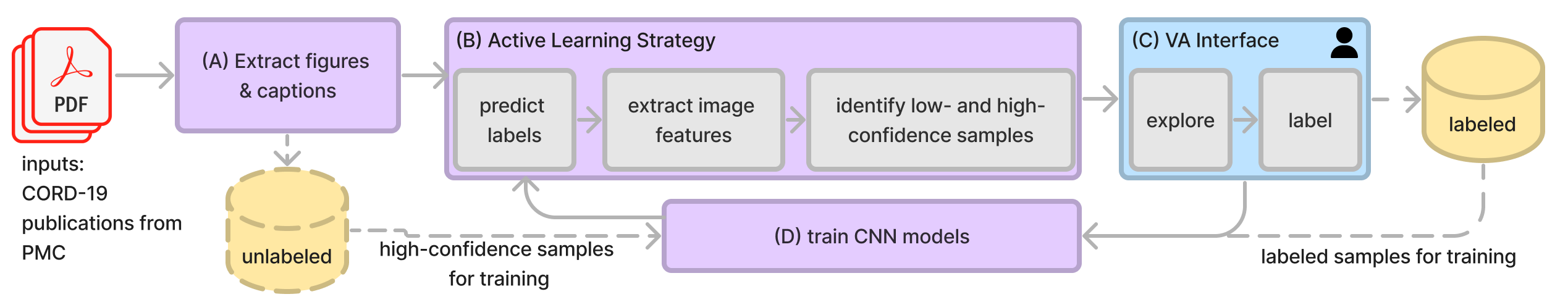}
  \caption{\label{fig:workflow}
  Architecture of BI-LAVA’s active learning system. (a) An offline pipeline extracts images and captions from PMC COVID-19 related publications. (b) The AL strategy uses previously trained CNN models to predict labels, and extracts image features from unlabeled images by fetching latent representations from the last convolutional layer. Next, it separates images based on the classifier’s confidence according to two popular metrics. (c) A user accesses the visual analytics interface to understand the collection of images and label them. (d) Labeled images and high-confidence unlabeled samples are used to retrain the image classifiers.
  }
\end{figure*}

Our solution uses an iterative workflow that progressively processes an unlabeled collection of images by leveraging trained image classifiers and user inputs (Fig.~\ref{fig:workflow}). Following the VIAL framework~\cite{bernard2018vial}, BI-LAVA has three outputs: data, model, and visual analytics. Users interact with labeled and unlabeled images to generate labeled data, which triggers the model output as a hierarchical set of image classifiers. Finally, the visual analytics component amplifies the knowledge output by enabling the understanding of the data and the model behavior.

To tackle the novel challenge of classifying image content at different levels of granularity, BI-LAVA uses a hierarchical image taxonomy. Unlike other systems that deal with a single classifier, we couple the taxonomy with a hierarchy of image classifiers where each classifier is a parent node in the taxonomy. To address the problem of incomplete or missing labels at different levels of granularity, BI-LAVA leverages a human-in-the-loop AL pipeline. This strategy allowed us to use each labeled sample, even when incomplete. For example, an image labeled as microscopy is used by the top classifier, while an image labeled as microscopy.fluorescence is used by the top classifier AND the microscopy classifier. Furthermore, this pipeline allows us to identify and correct issues in both the data and model. By allowing for an exploration of the dataset with the front-end VA module, BI-LAVA further addresses the challenges of working with unfamiliar datasets while improving data quality.

\subsection{Data abstraction}

Our system relies on data from images and the hierarchical information of the taxonomy of modalities. Most images were extracted from scientific publications from PubMed Central after we extracted and segmented subfigures. In addition to using the raw image data to train the classifiers and display thumbnails, the image schema includes the data source,  split set type (training, validation, test, unlabeled), and label and captions (although not available at subfigure level, ie., sub-caption), if available. We further derive latent representations, predictions, and associated probabilities using deep learning models. Other attributes include projected coordinates to 2D and metrics for the AL strategy (margin sampling~\cite{scheffer2001active}  and entropy~\cite{shannon2001mathematical}). Our taxonomy of modalities is a tree structure where each child node increments the category's detail. Also, each parent node is associated with an image classifier and its performance metrics.

\subsection{Activity and task analysis}

We formulated the activities and tasks from our experience designing a labeling interface for biocuration~\cite{trelles2021animo} and from our attempts to find datasets that matched the specified taxonomy. Our previous labeling interface allowed biocurators to label images within publications on a one-publication-at-a-time basis, imitating their curatorial process. Although our system reduced the labeling time, selecting publications to label made it challenging to identify under-represented samples, leading to imbalanced datasets. Our initial dataset required merging images gathered from selected publications, competitions~\cite{GSB2016}, synthetic charts~\cite{adobe-synt}, and other biomedical sources~\cite{demner2012design} in addition to a set of images labeled by curators. This laborious process was time-consuming because biomedical images required extra time to verify our labeling decisions. As model builders, we needed more domain knowledge to discern between modalities easily.

After the start of the COVID-19 pandemic, our collaborators wished to shift our image-harvesting efforts from biomedical papers to a growing collection of COVID-19 publications~\cite{wang2020cord19}. However, our manual labeling interface did not scale well to the much larger number of COVID-19 documents (more than 250,000). Our group could also not afford the time to manually inspect samples to validate labels and image quality. Instead, we aimed to leverage the image classifiers we had developed for the much smaller set of labeled images and leverage our experience as model builders to replace biocurators as data annotators. However, as we approached the new classification problem and researched possible solutions, we learned that other model builders faced similar problems across domains related to the same lack of domain expertise with a dataset. Consequently, providing visual cues to support data understanding before labeling became essential.

We documented the issues encountered by our group and problems reported by other engineers with similar issues, and conducted semi-formal interviews with several ML practitioners and with our biocurator collaborators. By further contrasting the resulting needs with tasks characterized for instance selection~\cite{Bernard2021Taxonomy} and supported by visual encodings~\cite{sarikaya2017scatterplots} (both discussed in the Related Work), we arrived at the following list of activities and tasks~\cite{marai2017activity}:

\textit{\textbf{A1}. Analyze labeled and unlabeled images for each class defined by the hierarchical taxonomy}
\begin{itemize}
    \itemsep0em 
    \item T1.1 Display an overview of the labeled samples and their relationship to the image models used in the classification taxonomy
    \item T1.2 Show images by similarity and allow the comparison between labeled and unlabeled data subsets
    \item T1.3 Explore local neighborhoods of images to become more familiar with data characteristics and characteristics shared between similar images
    \item T1.4 Identify how model behavior changes for different types of images, and what features commonly lead to labeling errors
\end{itemize}

\vspace{0.5mm}
\textit{\textbf{A2}. Select candidates to update and label images}
\begin{itemize}
    \itemsep0em 
    \item T2.1 Identify labeling errors in the labeled and unlabeled sets corresponding to each image classifier
    \item T2.2 Update image labels for single or multiple images
    \item T2.3 Browse low-confidence samples to identify labels that require human-intervention
    \item T2.4 Browse high-confidence samples to support the evaluation of the pseudo-labels
    \item T2.5 Evaluate modifications to data labels
\end{itemize}

Beyond general exploratory labeling (T1.1, T2.1~\cite{Bernard2021Taxonomy}), T1.2 - T1.4 are specified for our project's labeling needs, although they resemble tasks typically supported by scatterplots~\cite{sarikaya2017scatterplots}. T2.3 - T2.4 are unique to our domain problem.
~\cite{Bernard2021Taxonomy}
Non-functional requirements included handling large image datasets of up to 500,000  images while allowing an interactive browser experience and supporting on-demand calculations of dimensionality-reduced features (a computationally-expensive operation). Finally, our system should be easily usable for model builders with limited visual literacy.

\subsection{Pre-labeling and image classifiers}
\label{section:prelabeling}

Pre-labeling refers to annotating data offline before training the classifiers~\cite{bernard2018vial}. We obtained labeled data from a few labeled datasets, including ImageCLEF 2013 and 2015~\cite{GSB2016}, from approximately 6,000 data labeled by biocurators on our previous labeling interface~\cite{trelles2021animo}, and from a synthetic charts dataset~\cite{adobe-synt}. In addition, we scrapped around 2,000 images from Open-i~\cite{demner2012design} and added approximately 15,000 samples of \textit{experimental} images provided by our collaborators. Except for the synthetic dataset for charts, all the other images come from scientific publications that guarantee verifiable provenance to a biomedical source. We note that, in contrast, searching the web for ‘light microscopy’ returns generic images of microscopes instead of the expected light microscopy images.  Our training efforts focused on a subset of classifiers due to labeled data availability (green nodes in Fig.~\ref{fig:taxonomy}). We made large efforts to guarantee that the data was suitable for training; however, we could not guarantee that mislabeling did not happen. In addition, we disregarded samples labeled by biocurators that included extraction errors (e.g., wrong segmentations in Fig.~\ref{fig:case-study}g) generated while extracting content from PDF documents. We manually inspected the samples for every labeled dataset and matched them to our taxonomy. Our total pool of labeled images contained 333,998 samples.

We trained each parent node of the taxonomy as a supervised classifier independently where the classes were the node’s children. To keep the training sets consistent across parent and child classifiers, we first trained the lower levels of the hierarchy using a stratified partition with training, validation, and test partitions of 70/10/20. For the parent classifiers, we follow a similar approach but also force images previously selected for a training set to remain in the parent’s training set and avoid data leakage. For instance, samples in the training dataset of the \textit{electron microscopy} classifier also belonged to the training dataset of the \textit{microscopy} classifier. Given its competitive results, our chosen architecture was a ResNet18~\cite{he2016deep} model. We trained these models using transfer learning from ImageNet and early stopping to avoid overfitting, with a learning rate of $1e-4$, and saved the F1 scores.

Given our collaborators' requirements for gathering COVID-19-related images, we created our unlabeled dataset from the CORD19 dataset~\cite{wang2020cord19}. This dataset contains entries to research articles about COVID-19, updated periodically until recently. We extracted approximately 32,000 documents from that subset available before January 2021 from PubMed Central. As the CORD19 dataset does not provide images, we used the PMC identifiers to collect the publications as PDFs using NIH's interfaces. We detail the pre-processing steps next.


\subsection{Active learning strategy}

We obtained the biomedical images for the unlabeled dataset by pre-processing publications in PDF format. First, we pass every document through an extraction pipeline~\cite{trelles2021animo} to obtain every figure, subfigure, and corresponding caption (Fig.~\ref{fig:workflow}A). The pipeline internally invokes modules for image and captions extraction~\cite{li2018compound}, and figure separation~\cite{li2019figure}. We follow this approach as figures in biomedical publications can contain subfigures with different modalities that would add noise to the training data. We then store these images in our unlabeled database.

Next, we start our AL strategy on the unlabeled collection of images (Fig.~\ref{fig:workflow}B). Using the pre-trained models, we infer image labels and prediction probabilities and extract the image features (from the last convolutional layer in the model before the softmax layer). We save the image features for the dimensionality reduction step later in the workflow. This step is repeated per classifier in the taxonomy from top to bottom following a branch of the taxonomy in order to treat each modality independently. High-confidence samples are identified in the dataset as images with an entropy~\cite{shannon2001mathematical} below $<0.05$, based on the CEAL framework~\cite{wang2017ceal}, which are used as pseudo-labels. Images with lower confidence are ranked using the margin sampling score~\cite{scheffer2001active} to provide more information to prioritize labeling decisions.

The following step in the workflow involves a model builder interacting with our visual analytics interface (Fig.~\ref{fig:workflow}C). We describe the components in Sec.~\ref{sec:design}. In this stage, the model builder produces new labels by confirming or correcting predicted labels. Confirmed pseudo-labels become training data in the AL approach. Finally, the training step (Fig.~\ref{fig:workflow}D) uses labeled data and high-confidence samples as a training pool to retrain the models from scratch before reinvoking the AL steps. Because training the model takes a considerable amount of time, we only retrain the model when training pools for a classifier, including new or updated labeled samples, are incremented by a user-defined threshold.

\subsection{Front-end design}
\label{sec:design}

\begin{figure*}[tb]
 \centering
 \includegraphics[width=\linewidth]{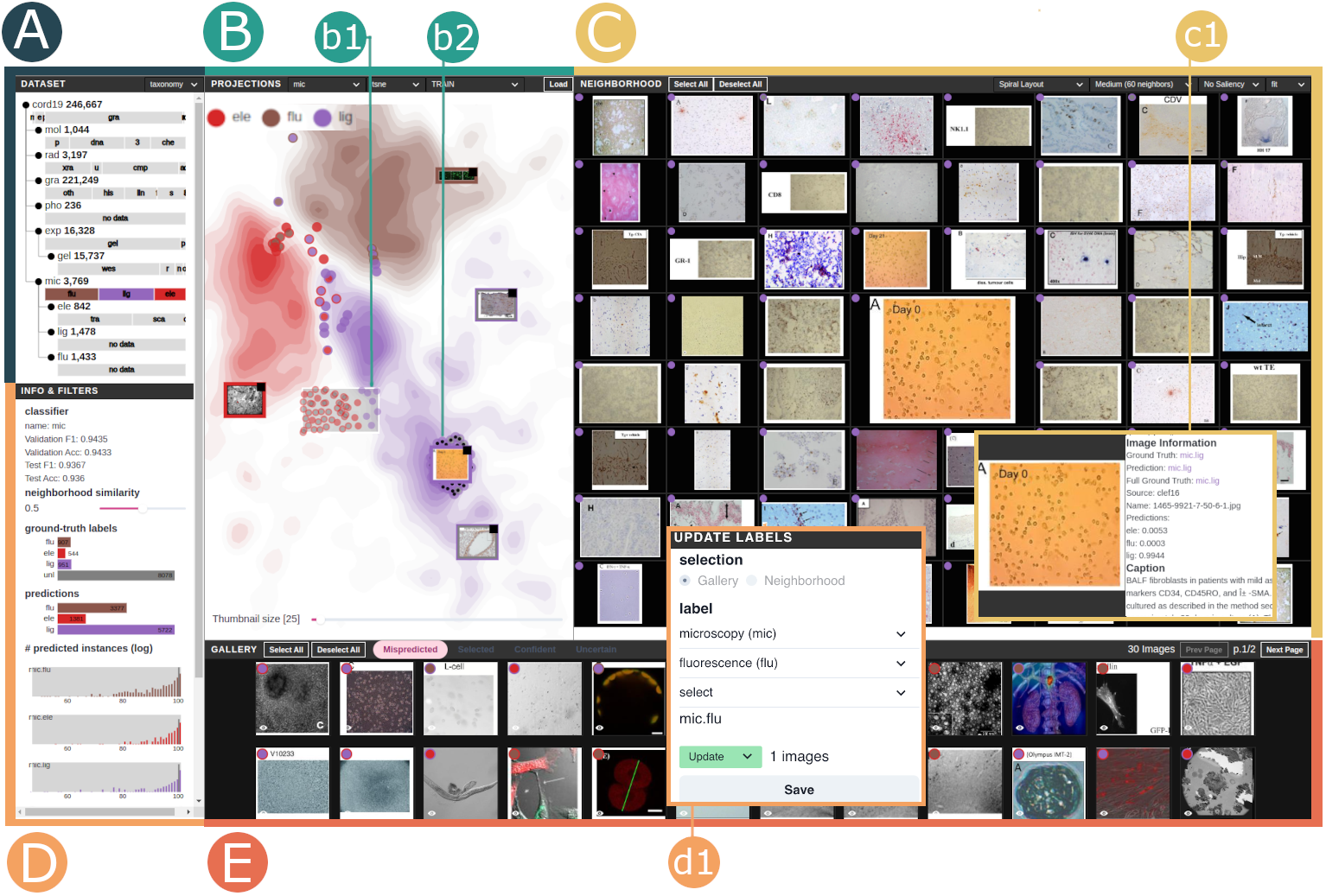}
 \caption{\label{fig:interface}
 Visual interactive labeling with BI-LAVA. (A) \textit{Dataset view} displaying a taxonomy of image modalities and proportion of samples per class. (B) \textit{Projection} of a training set of microscopy images using t-SNE. Density heatmaps hide confident samples, while circle encodings show samples with a neighborhood similarity below 50\%. Brushing reveals samples in the lower region (b1). (C) \textit{Neighborhood view} displaying an image of interest from the scatterplot (b2) and neighbors in a spiral layout. Details on demand show the image caption and prediction probabilities (c1). (D) Performance metrics from classifiers and filters. Scrolling down reveals the \textit{Update panel} (d1 positioned on the side for readability). (E) \textit{Gallery} of images for different data subsets: mispredicted samples, samples from the selected area, confusing samples, and uncertain samples.}
\end{figure*}

We designed our visual analytics platform to address the previously discussed challenges of model builders when dealing with unfamiliar datasets, ultimately leading to more efficient labeling. BI-LAVA (Fig.~\ref{fig:interface}) provides different views and interactions to support data understanding and labeling. The \textit{Dataset view} (A) provides an overview of the distribution of labeled samples per class in the taxonomy (\textit{T1.1}) and an overview of the current label updates (Fig.~\ref{fig:updates}) in a user session (\textit{T2.5}). The \textit{Projection view} (B) allows the exploration of projections from image features (\textit{T1.2-1.3, T2.1}) and inferring model behavior (\textit{T1.4}). Users can filter the data using the panel to its left (D). The \textit{Gallery view} (E) provides different entry points to images (\textit{T1.3,T2.3,T2.1,T2.4}). The \textit{Neighborhood view} (C) enables exploring the most similar thumbnails (\textit{T1.3-1.4}) and explains predictions (\textit{T1.4}). Finally, users can update labels (\textit{T2.2})(d1). This design results from a parallel prototyping approach~\cite{dow2010parallel} with feedback from three visual computing researchers. We implemented the front end using React, D3~\cite{bostock2011d3} and ThreeJS, and used Flask, PyTorch, RAPIDS, and MongoDB for the back end.

\subsubsection{Dataset view.}
\label{sec:dataset}

The \textit{Dataset view} provides an overview of the labeled data and labeling updates performed in a session. Visual cues attempt to guide model builders' exploration, such as showing classifiers with imbalanced distributions of labeled data. After selecting a node in the taxonomy from the top left of the interface, the view for labeled data summarizes the distribution of labeled images per category as an indented list (Fig.~\ref{fig:interface}A) with the total number of samples per category. In this tree-like structure, each node represents an image classifier. Below, a horizontal bar takes the whole available width space to display the distribution of images per category (\textit{T1.1}). Consequently, the bars are not comparable between classifiers. We made this design decision as our exploration focuses on one image classifier at a time. Bars are color-coded following a chromatic scale of 10 different colors. However, given the large number of nodes in the taxonomy, some categories may share the same color across different classifiers.

Design alternatives included trees and treemaps. However, juxtaposing the taxonomy and image distribution details provided a more compact and understandable representation. In contrast, treemaps produced small squares due to the imbalanced number of images between categories. Zoomable treemaps were another alternative, but they required more interactions and did not provide an overview of all nodes.

\begin{figure}[tb]
 \centering
 \includegraphics[width=\linewidth]{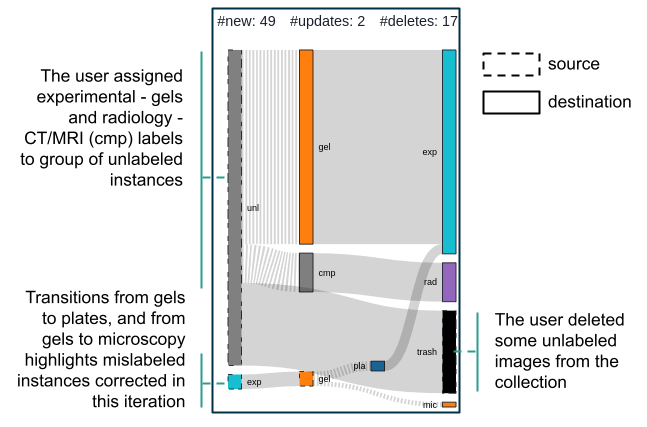}
 \caption{An alternative view in \textit{Dataset view} (Sec.~\ref{sec:dataset}) for displaying labeling updates during a user session.}
 \label{fig:updates}
\end{figure}

Estimating the number of labels to update before finishing a labeling session is a hard challenge. Few images may lead to no significant changes, while many label updates demand more participation from human annotators. Instead of recommending such a number, BI-LAVA displays an overview of the changes when the user selects the updates option in the toolbar. A Sankey diagram (\ref{fig:updates}) summarizes the label updates done so far in the labeling session, including new labels, updated labels, and deleted items (\textit{T2.5}). Boxes to the left (with dashed borders) represent the taxonomy categories affected by the updates, while boxes to the right (with solid borders) represent the updated values. A black box complements the boxes to the right as a placeholder for deleted elements. Our representation includes any parent category for an updated taxonomy. These update flows in the view provide cues for common mistakes in the current dataset, such as labeled samples moved to another distribution (i.e., mislabeled).

\subsubsection{Projection view.} 
\label{section:projection}
The \textit{Projection view} displays the images associated with an image classifier on a 2D scatterplot (Fig.~\ref{fig:interface} B). Due to the documented benefits for novice users, we followed the “overview first, zoom and filter, and details on demand” mantra~\cite{shneiderman2003eyes} to support data understanding. Toolbar options enable selecting the classifier, projection technique, and dataset sub-set. BI-LAVA calculates the dimensionality reduced features to 2D using either PCA, t-SNE~\cite{van2008visualizing} or UMAP~\cite{mcinnes2018umap}, leveraging the RAPIDS GPU-based implementation to obtain acceptable on-demand processing times (e.g, from 5.8s for 50,000 images to 3m for 500,000 images, using t-SNE). The dataset sub-set options include training, validation, test, unlabeled, unlabeled + train, or all the data. These options allow the user to narrow down the dataset based on their intent (\textit{T1.2-T1.4}). For example, training samples include ground-truth labels and high-confidence, which are more useful to get an initial idea of the data and whether data expected to be correct has labeling issues or erroneous predictions (\textit{T2.1}). In contrast, unlabeled + train shows how the unlabeled elements relate to the training pool.

We use a combination of a scatterplot and density heatmaps to represent the image data. Given the large dataset size, we first show an overview of the data by displaying the different clusters of images as density heatmaps. Similar to Splatterplot~\cite{mayorga2013splatterplot}, we then display images of interest on top of the heatmaps. Since we are interested in images that are more likely to be mislabeled (\textit{T2.1}), we show images whose neighborhood hit metric~\cite{paulovich2008least} is below a user-defined threshold. The rationale behind this decision is that users may be interested in exploring first images surrounded by different neighbors, which usually lie on cluster borders. Users can manipulate this threshold at will (Sec.~\ref{section:filters}). We encode each sample as a circle where the background color represents the ground truth, and the border color represents the prediction label, following the color scheme used in the \textit{Dataset view}. For unlabeled samples, we fill the circle background in gray. Image thumbnails appear as details on demand after clicking on a circle mark.

Analyzing these sub-sets enables understanding model performance and problematic regions~\cite{rauber2018projections}. Thus, this view depends on the notion of similarity obtained in the 2D space to guide exploration. However, these projections could suffer from distortion errors~\cite{nonato2018multidimensional}. The \textit{Neighborhood view} mitigates this effect by showing the raw image data (Fig.~\ref{fig:interface}C).

This view supports interactions to ease exploration and is linked to other views in the interface. First, users can zoom into an area with mispredicted samples to inspect neighbors and reveal their thumbnails by clicking on a circle mark. Then, clicking on the top right of the thumbnail triggers the \textit{Neighborhood view} (Sec.~\ref{section:neighbor}). Next, by brushing over a region, the view reveals the samples hidden due to the neighborhood hit threshold (Fig.~\ref{fig:interface} b1). This image selection is also displayed on the \textit{Gallery view} (Sec.~\ref{section:gallery}) as a grid layout of thumbnails. At last, the \textit{Filters view} allows filtering of the view by image metadata (Sec.~\ref{section:filters}).


\subsubsection{Filters view}
\label{section:filters}

BI-LAVA leverages image metadata to filter an image collection (\textit{T1.2}). Bar charts and histograms summarize the values of the current data sub-set, including unlabeled data if applicable. We use bar charts to show the distribution of the ground truth labels, predictions per class, and data sources. Each bar in the charts acts as a filter, and updates to the bar fill indicate the active filters. Users can identify discrepancies in model behavior by combining these filters. For example, filtering electron microscopy images should ideally display only images predicted as that category. Hence any other misprediction is worth reviewing.  
  
For the model predictions, we use a histogram for each class to show the log distribution of samples for each probability bin. We chose a log distribution to account for the imbalanced distribution of samples. Two color-coded bars are displayed side by side per prediction probability, one for labeled and one for unlabeled samples (\textit{T1.2}); this setup allowed us to gain vertical space. Sliders filter the values. For example, we can filter samples with prediction probabilities lower than 60\%. Finally, the top area displays the classifier’s performance metrics.

\subsubsection{Gallery view}
\label{section:gallery}
The \textit{Gallery view} (Fig.~\ref{fig:interface}E) complements data exploration by showing images of interest based on user interaction, classification outputs, and outcomes of the AL strategy. Content is organized using tabs. The first tab groups the mispredicted images. For example, discrepancies between ground truth and predicted labels indicate potential issues in the classifier (\textit{T2.1}). The second tab groups selected images from the \textit{Projection view}. Finally, the last two tabs group high-confidence samples, starting with the most confident images, and low-confidence samples, starting with the most uncertain images, respectively (\textit{T2.4},\textit{T2.3}).    

This view places these images using a paginated grid to show the image content. We place a circle mark for each image on the top left, following the same color-coding convention as in the \textit{Projection view}. In addition, a red circle mark on the top right indicates an image marked for deletion, while a color-coded circle mark indicates that the image label has been changed to the color-coded label. As images in our dataset have different aspect ratios, we chose squared elements to display and fit their content to the constrained space. However, hovering over a thumbnail expands the image content.

As an alternative to the "overview first" approach in the \textit{Projection View}, the gallery enables a "details-first, show context, overview last" approach~\cite{luciani2018details}. Model builders can click on an image in the gallery to reveal its context, depicted as a neighborhood of thumbnails (\textit{Neighborhood view}), and later inspect the neighborhood's location in the whole projection space. For instance, \textit{unknown unknowns}~\cite{chen2018anchorviz} appears when a sample with high confidence is mispredicted. If users spot one of these cases, they can inspect the neighborhood for more potential errors. Finally, selecting thumbnails indicate labeling candidates (Sec.~\ref{sec:label}).

\begin{figure*}[htb]
 \centering 
 \includegraphics[width=\linewidth]{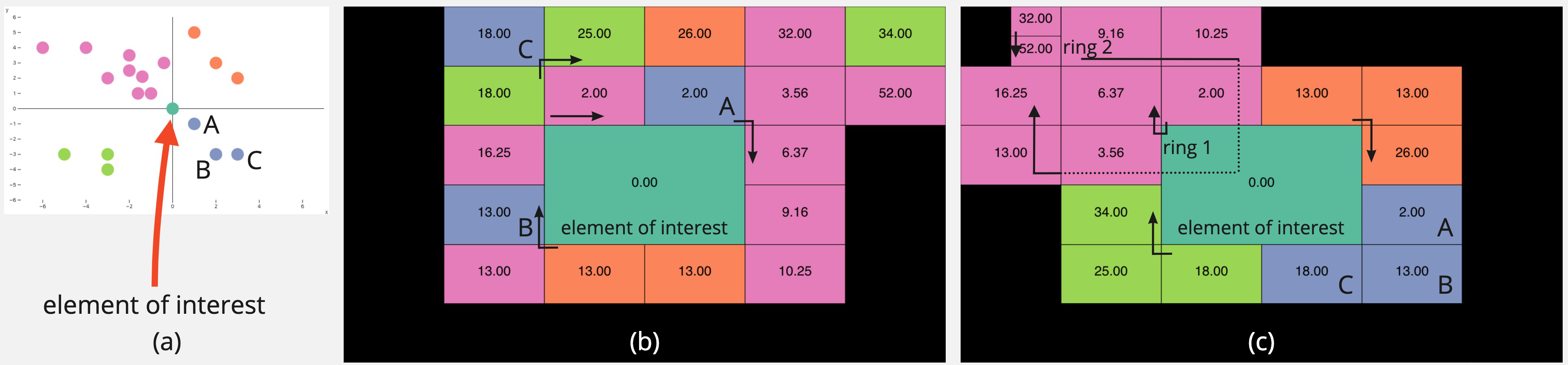}
 \caption{\textit{Spiral layout} vs \textit{Spatial spiral layout}.  (a) Random distribution of points in a 2D projected space, color-coded by Cartesian quadrant, surrounding a point of interest in the center. (b) Spiral layout with two concentric rings. Color encodes quadrants and unused space (black), and numbers show distances between an item and an element of interest. (c) Spatial spiral layout positioning points based on their quadrant and distance. The top-left quadrant requires subdivisions.}
 \label{fig:spiral}
\end{figure*}

\subsubsection{Neighborhood view}
\label{section:neighbor}

\begin{figure*}[tb]
 \centering
 \includegraphics[width=\linewidth]{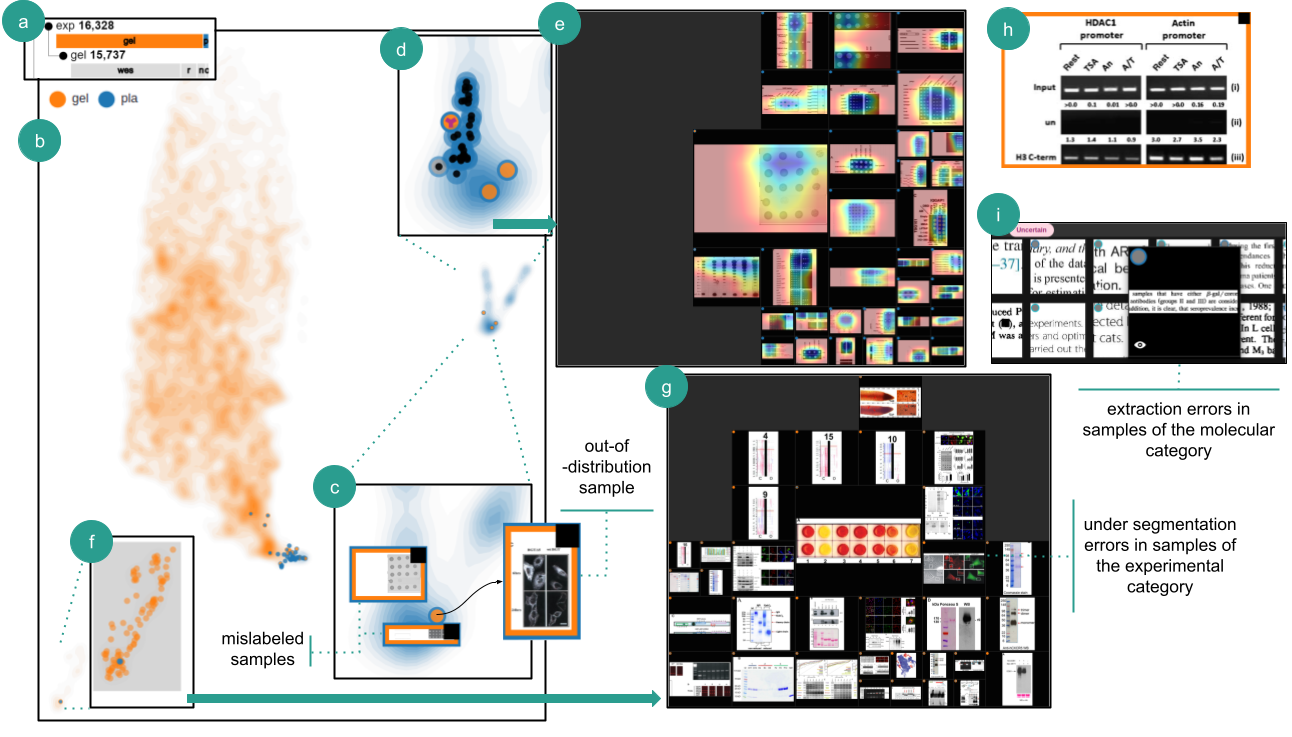}
 \caption{Interaction during the case study: COVID-19 Image Modality Labeling. (a) Imbalanced distribution of samples for gels and plates. (b) Projected image features for gel (orange) and plate (blue) samples using UMAP. (c) Zoom in to the plate region on the right displaying the thumbnails for three samples labeled as gel but predicted as plate. The cluster contains an out-of-distribution sample. (d) Mispredicted image of interest and immediate neighborhood in the plate cluster. (e) \textit{Neighborhood view} using a spatial spiral layout and saliency maps. Circular patterns yield higher activations (in blue). (f) A plate sample (blue) is predicted as gel (orange). (g) \textit{Neighborhood view} for a plate image with unconventional colors in a gel region. Neighboring elements include out-of-distribution and under-segmented samples. (h) An experimental gel sample. (i) Uncertain samples in the molecular category representing portions of text.}
 \label{fig:case-study}
\end{figure*}

While the projection view provides flexibility for viewing, filtering, and selecting regions in the dataset, visualizing image thumbnails is challenging due to clutter. We overcome this problem by inspecting image thumbnails on a separate view from the perspective of an image of interest. This rationale allows expanding the exploration from an image sample presented on the \textit{Gallery view} or found through investigation on the \textit{Projection view}. For example, an uncertain image can reveal more patterns in a neighborhood that may confuse the model. In particular, the \textit{Neighborhood view} (Fig.~\ref{fig:interface}C) sorts the neighboring images based on their proximity (i.e., similarity) to this image of interest using three layouts: a spiral layout~\cite{wang2015similarity} (Fig.~\ref{fig:interface}C), a novel variation of this spiral layout that preserves the relative position of the projected features in the 2D landscape (Fig.~\ref{fig:case-study}e,g), and a traditional grid layout. While grid layouts are more common than spiral layouts, we decided to use spirals as the default layout as they highlight the image of interest by placing it in the center of the canvas. The local exploration in this view aids the inspection of visual features per region. Understanding these similarities helps model builders create a mental model of the classifier's behavior (\textit{T1.3,T1.4}), like identifying regions with chest X-rays.

\textbf{Spiral layout.} The spiral layout for images~\cite{wang2015similarity} places the image of interest in the center and neighbors in concentric rings. The order of the neighbors, sorted by distance, starts from the top-left quadrant of the central image and proceeds clockwise (Fig.~\ref{fig:spiral}b). We calculate the neighbors based on the Euclidean distance in the 2D projected view and mitigate distortion errors by displaying the thumbnails in this spiral layout. In addition, the layout organization approximates the elements’ position in 2D to provide consistency. Each ring contains thumbnails with the same aspect ratio as the central image to fit perfectly on every edge of the central element. If required, the method is flexible enough to consider further size reduction of outer rings by dividing the width and height in multiples of 2. As the spiral layout follows a treemap algorithm, the spiral pattern can recursively appear for boxes with hierarchical elements. However, as image sizes shrink fast, the recursion affects the visibility of the thumbnails. Thus, we restrict our design to divide each rectangle into four elements at most if required (Fig.~\ref{fig:spiral}c). We use the spiral layout as the default setting.

\textbf{Spatial spiral layout.} Although the spiral map arranges elements by distance to the central image, the position of each element does not preserve the structure of the elements in the 2D projection. Figure~\ref{fig:spiral} displays an example case of this problem. The scatterplot (a) displays a set of random points, color-coded by the quadrant where they appeared, with the point of interest in the center. The spiral layout (b) uses the distance to the center to organize items in concentric rings. However, the resulting arrangement places items from the same quadrant far apart, violating the structure given by the scatterplot. Our alternative arrangement (c) organizes items by similarity while preserving quadrant membership. It should be noted that an unequal distribution of elements per quadrant requires a subdivision of the space to fit the boxes. These ideas lay the foundations for our \textit{spatial spiral layout}.

To build a spatial spiral layout, first, we choose the number of rings in the layout. Next, we build each ring by subdividing a sorted list of points by quadrant until we fill in all the rings. When elements in a new ring have the same size as in the previous ring, the new ring allocates four more elements. However, when the outer ring halves the size of the elements, it contains double the elements plus 4. Then, we traverse the list of ring elements in reverse order and subdivide each element into four. Our \textit{Neighborhood view} considers cases for 2, 3, 4, and 5 rings called small, medium, large, and very large views, respectively, that can allocate from 32 to 180 images (supp. materials).

\textbf{Features and interactions.} In contrast to the \textit{Gallery view}, scaled images match the rectangular space available in the layout. From previous experiences working with biomedical images~\cite{trelles2021animo}, images with unusual aspect ratios (e.g., skinny images) may indicate the presence of extraction errors while processing PDF documents. Showing thumbnails covering the squared space, as in the \textit{Gallery view}, is also available. Markers for labels and updated information are consistent with the design in \textit{Gallery view}. Additional features support the exploration and actions in the neighborhood of thumbnails. To provide a proxy for the interpretability of our image classifiers, we provide saliency maps using GradCAM~\cite{selvaraju2017grad}. We color-encoded the saliency maps using red for areas with low activations and blue for areas with high activations. Users can see more details by clicking on markers, such as captions, data sources, and predicted probabilities per class (Fig.~\ref{fig:interface} c1). Finally, as in the \textit{Gallery view}, users can indicate the images to update by clicking on the thumbnails or using the toolbar options.


\subsubsection{Update panel}
\label{sec:label}

This panel allows updating the selected samples from the \textit{Projection} or \textit{Gallery views} (\textit{T2.2}). Users can indicate new labels or mark items for deletion (Fig.~\ref{fig:interface} d1). We also allow specifying child labels or different parent nodes if the samples were misplaced (i.e., out-of-distribution).


\section{Evaluation}

Beyond manual labeling, which BI-LAVA partially automates and thus clearly outperforms, there are no other labeling systems that BI-LAVA could be reasonably compared quantitatively against. Instead, we evaluate BI-LAVA through two case studies with several practitioners and rigorous qualitative feedback. In the first case study, a senior model engineer with experience with biomedical images used BI-LAVA to analyze two subsets of images extracted from COVID-19 publications in conjunction with labeled sources. In the second case study, five ML practitioners independently used BI-LAVA on a subset of a biomedical image dataset, which was unfamiliar to them. In both case studies, we used a think-aloud protocol with note-taking, followed by a questionnaire.

\subsection{COVID image modality labeling}

This case study aimed to identify problematic instances from two types of imaging modalities in the COVID-19 dataset and update the labels accordingly. The study was completed by a senior model engineer working with the dataset as part of the biocuration project using a think-aloud protocol with note-taking. The engineer first sought to investigate the imbalanced distributions of samples he had used to train image classifiers—namely, the subset of gels and plates in the experimental (exp) assays category. Second, the engineer analyzed the molecular (mol) category, whose images are known for containing text strings such as RNA and DNA sequences. The experimental and molecular classes are handled by two different classifiers in the second level of the hierarchical taxonomy.

The engineer started his analysis by inspecting the \textit{Dataset view} to locate a modality category with an imbalanced distribution of samples (\textit{T1.1}). After noticing a significant disparity between the gel and plate sub-classes (15,737 gel vs. 591 plate samples) in the experimental node (Fig.~\ref{fig:case-study}a), he loaded the training and high-confidence unlabeled data using UMAP, due to previous experience with this technique (\textit{T1.2}). Figure~\ref{fig:case-study}b shows one big cluster in the center dominated by gels (in orange) and two small clusters on the sides, mainly grouping samples of the same class (plates shown in blue). He preferred first to analyze the small clusters to determine the reason for the presence of few gels in a dominant plate region, or vice-versa.

By zooming in on the cluster on the far right (Fig.~\ref{fig:case-study}c), the engineer identified three samples with a ground-truth label of gel predicted as plates within a blue region (plates). Then, he clicked on the instances in the \textit{Projection view} to visualize the thumbnails: two looked pretty similar, while the last one looked like an out-of-distribution sample (\textit{T2.1}). Next, the engineer tried to brush over the area to open more thumbnails, but the view became cluttered. Therefore, he selected one suspicious sample and opened the \textit{Neighborhood view} (Fig.~\ref{fig:case-study}d, ~\textit{T1.3}). As most neighbors appear to the sample's right, he switched the view to the spatial spiral layout (Fig.~\ref{fig:case-study}e). \textit{"It becomes clear now that this image is mislabeled as gel while it is a plate. They all share these circular patterns"}, he added. An inspection of the saliency map further corroborated that the circular patterns caused the highest activations in the model. The engineer then selected the other mispredicted sample, also a mislabeled instance, and updated the ground truth label to plate. The engineer commented on the remaining mispredicted gel sample: \textit{"I have seen these images before when working with the microscopy classifier. It probably is a fluorescence sample, so it is peculiar to find it classified as a gel"} (Fig.~\ref{fig:case-study}c right). After opening the details for each of the three images of interest and checking the image source, he added: \textit{"These three samples came from the same labeled collection we generated using a previous version of a classifier. I now see that we most likely got something wrong and need to revisit several labeled samples."} In addition, he identified several thumbnails with very narrow aspect ratios, which he confirmed were over-cropped figures that needed to be discarded (\textit{T2.1}). He ended the cluster inspection by deleting the errors and moving the mislabeled sample to the microscopy set (\textit{T2.2}).

Next, the engineer wished to check the isolated plate instance within the gel cluster on the left side (Fig.~\ref{fig:case-study}f). This sample looked like a plate but used a different color scheme (Fig.~\ref{fig:case-study}g). After brushing over the plate cluster previously explored, he hovered over the \textit{selected} tab in the \textit{Gallery view}. He commented: \textit{“For me, this image is a plate, but the color is different; no caption is available to be sure. Also, this gel cluster has many instances that should not be under the experimental category.”} After filtering samples by source, he stated again that one of the labeled datasets needs to be revisited as it contains noisy samples, including images with multiple modalities (\textit{T2.1)}.

Then, the engineer focused on the remaining mispredictions and evaluated the predictions in the unlabeled dataset. He noticed that the issues appear in a tiny area in the bigger cluster's boundaries of gels and plates. Thus, he decided to open the \textit{mispredicted} tab in the \textit{Gallery view} to understand the issues in these 58 instances (\textit{T2.1}). He quickly identified more microscopy images that should not have been in this dataset. Then, curious about the classifier's performance on the unlabeled dataset, he inspected the \textit{confident} tab (\textit{T2.4}). Most unlabeled samples were predicted as gels, and relatively few instances had issues. In contrast, plate predictions contained several out-of-distribution samples, such as radiology CT scans of chests and brains. He commented that these particular samples were problematic as the model classifies them as highly confident predictions. Based on this inspection of the confident samples, he decided to show only unlabeled samples on the \textit{Projection view}, brush over the dense gel regions in the middle, and open one unlabeled sample in the \textit{Neighborhood view}. The image caption listed the term \textit{western blot}, a subclass of gel. As all the images in the neighborhood looked quite similar, he displayed 180 neighbors and confirmed the label of the unlabeled samples as gel (\textit{T2.2}). He finalized this inspection by mentioning that the 99\% F1 score of the experimental classifier cannot be fully trusted as it hides many errors, such as mislabels and lack of representation.

Last, the engineer targeted the molecular dataset. He hypothesized that images containing text annotations could affect predictions for subclasses mainly containing characters, such as protein sequences. He started the analysis by plotting the unlabeled subset for the molecular (mol) category. Surprisingly, the \textit{uncertain} tab did not show instances of confusing biomedical images containing text but instead showed small chunks of text extracted from paragraph regions in the PDF (Fig.~\ref{fig:case-study}i). He marked these images for deletion and commented: \textit{“These samples indicate an error on our extraction algorithm. It would be good to mark them as an ‘other’ class within the molecular dataset as a proxy for these mistakes”}. In contrast, an inspection of the confident samples showed a good performance on chemical structures and DNA sequences, which appear as phylogenetic trees of nucleotide samples.


\subsection{ML practitioners exploring unfamiliar data}

This case study aimed to collect qualitative feedback about the benefits of BI-LAVA when an ML practitioner targets an unfamiliar dataset for data labeling. We recruited five ML practitioners to use BI-LAVA and perform a series of tasks to achieve this goal. Four out of the five practitioners had no previous experience with data visualization. Also, four practitioners were not affiliated with our project. Although we recognize the fifth practitioner as a co-author, his contributions focused on the AL components. We completed four in-person sessions and conducted the last session over Zoom and Parsec due to our remote collaboration. For in-person sessions, practitioners interacted with 31’’ monitors (1920x1080). For the remote session, the practitioner used a 15’’ monitor (1366x768). Every session lasted approximately 90 minutes and employed a think-aloud approach with note-taking.

At the beginning of each session, the practitioners read a description of the project background, and a facilitator addressed any additional questions. Then, the facilitator demonstrated the system features in one of the taxonomy classes. Next, the practitioners performed the following four tasks on the experimental assay (exp) dataset: (1) explore the dataset and identify characteristics for each class; (2) identify noisy samples; (3) identify high confidence samples from the unlabeled dataset to leverage as training samples; and (4) label low confidence samples from the unlabeled dataset. Finally, they provided qualitative feedback in an online questionnaire. During the session, the facilitator addressed any concerns. We chose the experimental assay as the unfamiliar dataset as it only contained two classes, and as shown in the previous case study, it had an imbalanced distribution of samples.

The first task aimed to familiarize practitioners with the dataset by exploring the samples used in the training dataset and the unlabeled samples with high confidence. t-SNE was the dimensionality reduction algorithm most used (4/5), in contrast to PCA and UMAP (1/5). Different exploration strategies (\textit{T1.2)} arose for this task, including sampling thumbnails across regions (similar to Fig.~\ref{fig:interface}B), exploring regions far from the overlapping region, starting from an overlapping region, and starting with low-density regions. During this stage, few practitioners used filters and preferred using the \textit{Neighborhood view} (\textit{T1.3}).

The most common approach to validate their hypothesis about the data was using the \textit{Neighborhood view} to inspect neighboring thumbnails. Some practitioners also used a combination of the \textit{Neighborhood}, \textit{Gallery}, and \textit{Projection views} by showing thumbnails in the scatterplot, selecting a distant region to display thumbnails on the gallery, and then contrasting those samples with the neighborhood. The spatial spiral layout was often preferred as it matched better the structure in the scatterplot. By the end of this task, practitioners noticed that circular patterns were prominent on plates (Fig.~\ref{fig:case-study}e) while stripes were more common on gels (Fig.~\ref{fig:case-study}h). A practitioner added that there was some grid organization between the elements in gels and plates.

For identifying noisy samples in the second task, practitioners followed different strategies (\textit{T2.1}). Most practitioners navigated to the overlapping regions or sought isolated samples on opposite clusters. For example, they identified a plate sample surrounded by gels (Fig.~\ref{fig:case-study}c). Another practitioner preferred the \textit{mispredictions} tab as a starting point for exploration, while one practitioner used filters to show gel labels predicted as plates and samples with low probability scores. Once the practitioners got more confident about the data, they commented on their desire to include functionality to filter out high-confidence pseudo-labels from the \textit{Neighborhood view} as labeled neighbors increased their trust compared to unlabeled ones. Saliency maps helped validate cases when the image of interest shared patterns with the neighbors and was mislabeled (Fig.~\ref{fig:case-study}e), but these heatmaps did not help explain well out-of-distribution samples (Fig.~\ref{fig:case-study}c).  

To identify the high-confidence samples to leverage during training, most practitioners remained in the \textit{training} subset and used the \textit{confident} tab (\textit{T2.4}). After iterating over that panel, three practitioners spotted out-of-distribution samples for the minority class (gels). Another practitioner preferred to inspect the unlabeled subset alone and use filters to visualize the regions for high-confidence samples (\textit{T1.2}). He explained that the lack of filters in the \textit{Gallery view} was a disadvantage; thus, he preferred to use the scatterplot. He added: \textit{“Predicted gels look good, but plates are a bit weird; it looks like there are not many predicted plates in this unlabeled collection.”} To inspect the low-confidence samples for the last task, they preferred to use the \textit{uncertain} tab (\textit{T2.3}). A couple of practitioners commented that they needed a way to look back at labeled data as they forgot some data characteristics. In particular, one of them iterated back and forth between projections. While none of the practitioners had major troubles updating labels, one indicated his preference for selecting elements per \textit{Neighborhood} rather than the sorted thumbnails in the \textit{Gallery view}.


\subsection{Feedback}

BI-LAVA yielded excellent feedback from the ML practitioners participating in our evaluation. For example, one practitioner told us he has to deal with many videos in his daily work; he commented: \textit{“(BI-LAVA) is so powerful to help a practitioner understand their data and develop ideas to improve the model. By the way, I like this tool so much, and if possible, I will consider using it in my research”}. In addition, practitioners agreed that the interface provided encodings and interactions helpful in speeding up labeling, identifying errors in labeled data, exploring and understanding the data in different ways, identifying imbalanced distributions, inspecting the quality of the pseudo-labels on high-confidence samples, and labeling low-confidence samples.

Although our system uses multiple views and entry points, practitioners commended BI-LAVA's ease of use. \textit{“Fairly intuitive and easy to use”}, one practitioner commented. \textit{"Good looking front-end, it's easy to use, and it has powerful functions"}, another added. Positive feedback was also given for these multiple views, filters, and supporting encodings: \textit{"(I liked) the filtering and neighborhood configuration, as well as the view that filtered images by different metrics. GradCAM was useful in figuring out what the classifier looked at for the more simple images and finding the noise's source"}. Regarding potential improvements, two practitioners suggested adding flexibility to the panels, such as readjusting their size or showing them on demand.

The interaction between the \textit{Projection} and \textit{Neighborhood Views} was the primary component to explore, understand, and validate the data. In addition, our spatial spiral layout served as a proxy for the projected image features: \textit{“... I was mainly interested in the neighborhood as a proxy for similar points, and a way to get more details on them, and less in terms of ranked similarity to a given image”}, commented one ML practitioner. Three practitioners also stated their preference for our spatial spiral encoding, highlighting its consistency with the projected samples; one preferred any spiral variation, and one stated he could use any layout. Regarding labeling, practitioners liked the multi-instance selection feature; a practitioner commented \textit{“having as many images as possible on the page is best”}.

Last but not least, one practitioner commented: \textit{“As a junior researcher and active competitor in ML contests, I spend more time designing models than visualizing datasets. I usually do error analysis of the top mispredictions to understand why my model might fail. I usually use Python and slowly interact with Jupyter. But, while using the interface, I noticed that this process can be improved and accelerated.”}


\section{Discussion}

Our evaluation shows that BI-LAVA aids in understanding the characteristics of different classes in a hierarchy, validating the quality of data and models, and labeling an unlabeled collection of biomedical images. Although BI-LAVA's multi-view interface is complex at first glance, our evaluation shows that ML practitioners could use it properly after only a few minutes. Furthermore, an expert audience adapted quickly to the interface complexity and provided positive feedback on usability.

Results from our case study with a domain expert highlight the importance of reviewing data quality for AL. The \textit{Projection view} helped the user spot mislabeled images. The target exploration in the \textit{Gallery} view led to identifying mistakes in the extraction pipeline. For all cases, the \textit{Neighborhood} view was essential to compare images and expand the search for similar images, which supports our multi-view design. Identifying errors through our interface supported the evaluation of his extraction and modeling tools on a broader data scope.

Results from our second case study with ML practitioners demonstrate BI-LAVA's potential to familiarize users with an unfamiliar dataset. Our multiple-classifier strategy helped users focus on particular data subsets while BI-LAVA leveraged incomplete ground-truth labels to provide information in the form of latent features and predictions. In addition, through multiple exploration strategies involving single and multiple views, users could explain the characteristics of each data class, identify and correct mislabeled instances and even reason about issues in parent classifiers that inserted out-of-distribution samples to the inspected subset.

The literature on visual interactive labeling formalizes several strategies for single-view exploration~\cite{Bernard2021Taxonomy}. We confirm these strategies are also used in our \textit{Projection} view. For instance, users explored the projected samples by examining areas with low and high density first, inspecting overlapping regions first, checking elements with disagreement among neighbors, or sampling thumbnails among areas for coverage.

While a single view can support many labeling strategies~\cite{Bernard2021Taxonomy}, our work shows that multiple connected views further expand these actions. Exploration following an overview-first approach~\cite{shneiderman2003eyes} benefited the practitioners with no previous experience with our dataset as it allowed them to leverage the spatial positions on the layout or data characteristics through filters to get a better grasp of the data. Once users got more familiar with the dataset, the details-first approach~\cite{luciani2018details} through the gallery became more widely used. In particular, users started to use the feedback from the AL backend when inspecting low- and high-confidence samples. The preference for displaying thumbnails in spiral layouts in contrast to a grid layout suggests the benefits of maintaining the spatial similarity between items. Our spatial spiral layout matched the projected view more closely than the spiral layout. In addition, we observed that displaying thumbnails in a juxtaposed view, as opposed to a superimposed one, encouraged more comparisons between regions, as users can visualize more images simultaneously without clutter. 

Our second case study also suggests that the user guidance provided by our neighborhood view is helpful when exploring projected data and avoiding misleading exploration strategies. In particular, after projecting the data using t-SNE, one user explored first groups of images further from the overlapping areas between class boundaries. Then, progressively the user looked at groups closer to the overlapping region while checking the variations among image groups. The rationale was that groups closer to an overlapping area might have similar features to another image class. However, the distortions resulting from t-SNE may provide incorrect cues for the t-SNE projection used. As Wattenberg et al.~\cite{wattenberg2016how} mentioned, distances may not map to the user's intuition and are tightly coupled to the perplexity parameter, which we did not allow users to manipulate. Enhancing our \textit{Projection view} with encodings to show distortions can avoid these wrong conclusions~\cite{nonato2018multidimensional}. We minimize these problems by using the \textit{Neighborhood View} as a secondary layout enrichment~\cite{nonato2018multidimensional}, which requires additional interaction from the user. 

Reflecting upon this experience, we identified the following three lessons for researchers addressing similar challenges:
\textit{L1. Use complex multiple views, as long as they support desired functionality.} Although conventional wisdom recommends simple interfaces for novice users, we found that our clients were not deterred by the multiple coordinated views. In fact, they appreciated the multiple data views, perhaps because the system provided a much desired functionality.

\noindent\textit{L2. Show neighborhood information to support trustworthiness.} Our clients were less concerned about precise similarity ranking, and more concerned with identifying data trustworthy samples. Preserving neighborhood spatiality (based on dimensionally reduced embeddings) was beneficial in this context, as was using connected views to reveal thumbnails. However, researchers should be aware of projection distortions and inform users about them.

\noindent\textit{L3. Provide slicing support to find and contrast trusted items.} We noted that clients who were unfamiliar with the data tended to rely on finding samples they could trust as anchor points. Allowing contrasting data slices (e.g., train/validation/test sets) between and against unlabeled sources helped when evaluating the generalizability of the models and data quality. 

Regarding limitations, BI-LAVA assumes that model builders can become familiar with a new dataset through progressive exploration. However, our evaluation used classifiers trained beforehand. Thus, we did not consider more complex bootstrapping scenarios due to the lack of representative samples for some subclasses. Next, given our goal of providing an efficient labeling tool, we focused on the functionality of providing multiple views, not on gathering metrics to compare different layouts. Defining tasks or metrics to compare the three layouts for visualizing thumbnails remains beyond the scope of this work. As BI-LAVA’s front end only depends on algorithms to classify low and high-probability samples, more recent AL or other strategies providing the same type of low and high samples could easily replace the CEAL component. Model training can be further expanded to use self-supervised learning in an AL setting~\cite{margatina2021al} to improve the use of unlabeled data. Finally, GradCAM might fail to produce correct explanations and can fail to adversarial attacks~\cite{viering2019how}, where appropriate. Our support for neighborhood exploration, however, helps to alleviate this issue by allowing users to explore additional evidence.   

BILAVA is designed for scalability. Our \textit{Neighborhood view} displays and enables labeling up to 180 images at once, and we have been able to render up to 850,000 images simultaneously in the \textit{Projection view}. The color and shape patterns support pre-attentive similarity detection~\cite{marai2018precision}, even for images on the spiral periphery, which may be less legible. Zooming, filtering, and visual summaries help navigate the data effectively. However, the \textit{Gallery} view's design, supported by our AL strategy, attempts to show prioritized elements to identify regions of interest with images of low and high confidence. Thus, this view does not attempt to scale to avoid a cognitive overload. Similarly, our dataset view relies on aggregated data but is limited in the number of classes shown at each node in the taxonomy.

Although we initially developed BI-LAVA to label biomedical images, it can generalize to other image classification problems in a semi-supervised setting by updating the input classification taxonomy, which can also be non hierarchical. BI-LAVA's methodology could also be adapted to non-image classification problems using a suitable visual representation of the data items to replace the image thumbnails. Future work directions include further generalizability to other domains by enabling visual editing of taxonomies, supporting parametrization of the dimensionality reduction algorithms~\cite{chatzimparmpas2020tvisne}, and providing a visual summary of the changes in model performance across iterations. 

\section{Conclusion}

In this work, we described the design and evaluation of BI-LAVA, a novel and timely labeling system initially developed for the hierarchical labeling of image modalities in biocuration. BI-LAVA integrates a visual analytics interface and an ML strategy for deep image classifiers to help non-experts understand a biomedical dataset, correct data quality issues, label samples, and reason about model behavior. In addition, BI-LAVA uses multiple classifiers to support a hierarchical taxonomy that deals with incomplete ground-truth labels. We described the characteristics of these biomedical images and identified the requirements based on a long-term collaboration with biocurators and text-mining researchers in the biocuration domain. Furthermore, we introduced custom views that support the understanding and labeling of unfamiliar datasets and exploring neighborhoods of thumbnails leveraging spatial constraints. Finally, our evaluation with machine learning practitioners and collaborators proves the usefulness of BI-LAVA, with a particular interest in the familiarization of non-experts with an image dataset.


\section*{Supplementary Materials}
Video demonstration, additional layouts, and spiral layout code are available at \url{https://osf.io/nvbtp/?view_only=d9aea97c77944916abadd8e65f9eee2f}

\section*{Acknowledgements}
We thank all members of the Electronic Visualization Laboratory for their feedback and technical support. We also thank the Research Experience for Peruvian Undergraduates (REPU) program for supporting William Berrios. We acknowledge awards from the U.S. National Institutes of Health (NLM R01LM012527, NCI R01CA258827) and the U.S. National Science Foundation (CNS-1828265, CDSE-1854815). We dedicate this work to the memory of Dr. Hagit Shatkay.

\bibliographystyle{eg-alpha} 
\bibliography{egbibsample}       
\end{document}